\documentclass[12pt]{article}
\usepackage[a4paper, margin=0.7in]{geometry}

\usepackage{amssymb}
\usepackage{amsmath}
\usepackage{soul}

\usepackage{float}
\usepackage{graphicx}
\usepackage[sorting=none]{biblatex} 
\begin{document}

\newcommand{\ket}[1]{\left | #1 \right \rangle}
\newcommand{\state}[1]{\left | #1 \right \rangle}
\newcommand{\bstate}[1]{\left \langle #1 \right |}
\newcommand{\bra}[1]{\left \langle #1 \right |}

\title{Mechanically Designing Protected Superconducting Qubits}
\author{Trevor McCourt}
\date{March 29, 2022}

\maketitle

\begin{abstract}
    Significant progress is required in the engineering of large, interacting quantum systems in order to realize the promises of gate-model quantum computing. Designing such systems is challenging, as the dynamics of continuous variable quantum systems are generally unintuitive, and brute-force numerical solutions are difficult to impossible in more than a few dimensions. In this work, I draw analogies between modern superconducting qubits and mechanical mass-spring systems in attempt to gain a simple intuition for what makes each design special. In particular, I analyze superconducting qubits that are inherintly protected from noise, and connect this protection to features of the corresponding mechanical system. The hope is that intuition gained from analyzing these systems mechanically will allow for intuitive design of useful superconducting circuits in the future. 
\end{abstract}

\section{Introduction}

Quantum mechanics most accurately predicts how the physical world behaves on the smallest scales. For example, cutting edge quantum electrodynamics calculations of the fine structure constant, which governs the strength of interaction between fundamental charged particles, agree with experimental measurements to less than a part-per-billion \cite{odom06}. It is therefore desirable to develop improved tools for studying it's implications. While computing technology based on the logical bit \cite{shannon48} has undoubtedly improved life over the last century, it is hopelessly inefficient at simulating generic quantum mechanical systems, and therefore the natural world. This motivates the development of a qubit, an abstract two level quantum system that is completely controllable and behaves perfectly unitarily. A large system of interacting qubits could be used could be used to effeciently emulate some natural system of interest. This was the founding premise of quantum computing \cite{feynman1982simulating}. Additionally, the field of quantum information \cite{nielsen00} has risen and found some classical problems in which mapping the problem to a quantum system yields some advantage \cite{Shor_1997}. 

Realizing such qubits is extremely challenging. Any quantum system that can actually be built in a lab generally does not behave like a perfect qubit, and cannot be controlled perfectly. Real qubits couple to the surrounding environment, which is uncontrolled, leading to a loss of control over the qubit state. The solution to this is quantum error correction \cite{Calderbank_1996}, which generally attempts to encode $M$ perfect qubits in $N > M$ physical qubits. However, the number of physical qubits required per logical qubit, and therefore the engineering burden of building a system based on quantum error correction, tends to grow rapidly as physical qubit quality decreases \cite{Fowler_2012}. Therefore, it is desirable to make physical qubits as good as possible before attempting to build an error correction system around them. This generally involves in trying to encode qubits in so-called dechoherence free subspaces of physical systems, which are parts of a physical system that do not easily couple to their surroundings. 

Physical qubits can be realized using a number of different quantum systems, such as trapped ions, neutral atoms, photons, spins, and superconducting electric circuits. Each has it's merits. In particular, atomic qubits generally take advantage of selection rules between electronic states to encode qubits in states that couple very weakly to the environment. Atomic qubits can stay decoupled from the enviromnet (coherent) for seconds at a time \cite{Ruster_2016}. The story is similar for photonic qubits. The tradeoff is that these states are fundamentally equally hard to influence via external controls, meaning that logical operation times tend to scale with coherence time; there is no free lunch to be had from systems given to us by nature. Engineering quantum systems from the ground up presents a possible escape from this proportionality. In essence, if we are the designers of a quantum system we may build in a back door inaccessible to simple natural forces that allows us to rapidly modify the system state while it remains protected from noise. This is one of the general goals of modern superconducting qubit design, which will be the topic of this work.

\begin{figure}
    \centering
    \includegraphics[width=\textwidth]{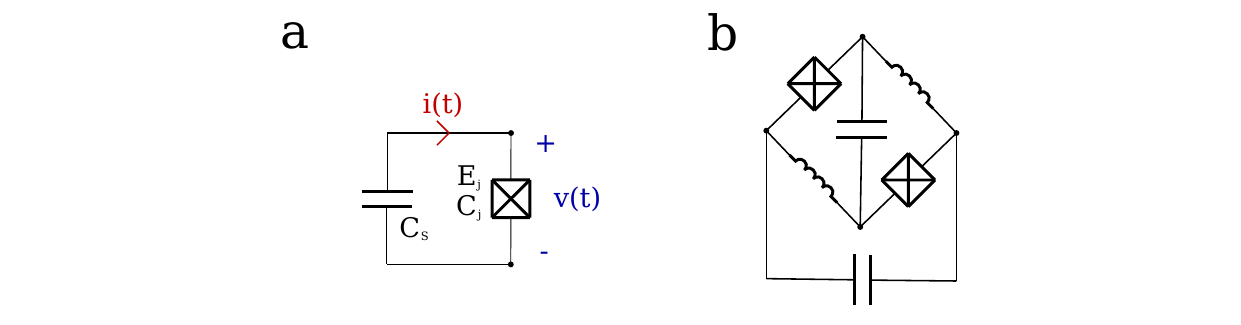}
    \caption{\textbf{Circuits including the Josephson junction a)} A capacitively shunted Josephson junction. $E_j$ is the Josephson energy of the junction and $C_j$ is the parasitic capacitance. $C_S$ is some shunting capacitance that combines in parallel with $C_j$ to produce the total capacitance $C$. \textbf{b)} A more complicated Josephson junction circuit.}
    \label{fig:cap_s_jj}
\end{figure}

Superconducting qubits are based on the Josephson Junction, which is a non-linear electric circuit element with the constitutive current-voltage relation \cite{Vool2017},
\begin{equation}
    i(t) \propto \sin{\left(\frac{\phi(t)}{\Phi_0}\right)}
\end{equation}
Where $\phi(t)$ is a generalzied flux, the integral of the voltage across the junction over all time,
\begin{equation}
    \phi(t) = \int_{-\infty}^{t} v(t') dt' 
\end{equation}
And $\Phi_0$ is the magnetic flux quantum. The Hamiltonian for a junction characterized by it's Josephson energy $E_j$ shunted by a capacitance $C$, as in Fig.~\ref{fig:cap_s_jj} a is therefore \cite{Burkard2005},
\begin{equation}\label{eqn:cap_jj_ham}
    H = \frac{q^2}{2 C} + E_j\left(1 -  \cos{\left( \frac{\phi}{\phi_0}\right)}\right)
\end{equation}

Where $q$ represents the charge on the capacitor plates, canonically conjugate to $\phi$, $\{q, \phi\}=0$ and in the quantum case $[q, \phi] = i \hbar$. This is similar to the Hamiltonian for the traditional harmonic oscillator, with the harmonic potential replaced by the cosine potential. The harmonic oscillator has equally spaced energy levels, and therefore does not straightforwardly implement a qubit, as the energy levels are not uniquely addressable. The cosine potential breaks this equal spacing, and for example the lowest two levels of such a circuit may be used to encode a qubit. This is the operating principle of the cooper-pair box, the first superconducting qubit to be realized \cite{shnirman97}. 

Modern superconducting qubits involve more complicated circuits, such as the circuit shown in Fig.~\ref{fig:cap_s_jj} b. This is the circuit of the $0-\pi$ qubit, a type of protected qubit \cite{Brooks2013}. This circuit has 4 total nodes which implies 3 degrees of freedom. Therefore, exactly finding it's energy levels will correspond to solving a continuous variable Schrodinger equation in three dimensions, which is nearing the limits of simple exact computational schemes (solving larger problems exactly may require a quantum computer!). It is relatively easy to dream up different circuts involving Josephson junctions, but extremely difficult to evaluate if the circuit is suitable for implementing a qubit, and more difficult still to tell if that circuit will have nice properties, such as protection from noise and easy control. This work will attempt to make use of a mapping between the Hamiltonian of such a circuit and simple mechanical systems, which may allow for increased intuition and ease the design process of such devices in the future.

\section{Mechanical Systems with Cosine Potentials}

\begin{figure}
    \centering
    \includegraphics[width=\textwidth]{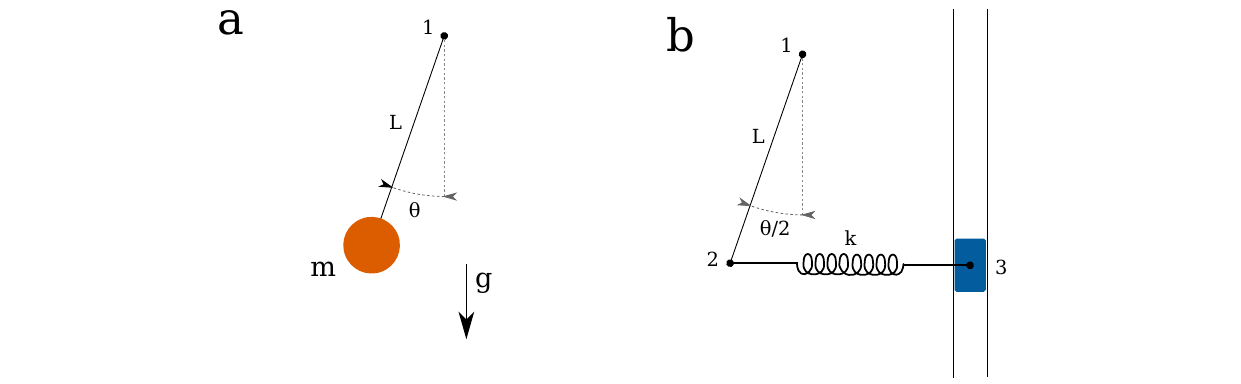}
    \caption{\textbf{Mechanical systems with cosine potentials} Two examples of mechanical systems with potential energies that vary as the cosine of the relevant coordinate, analogous to the potential of a Josephson Junction. \textbf{a)} A pendulum of mass m and length L in a vertical gravitational field. The pendulum is pinned at (1), and is free to rotate about this point. \textbf{b)} A rotating rod of length L connected to a spring pinned to a massless and frictionless slider, which ensures the spring remains horizontally oriented regardless of $\theta$. The rod is pinned at (1), and the spring is pinned at both (2) and (3). }
    \label{fig:cos_pot}
\end{figure}

While the cosine potential present in equation \ref{eqn:cap_jj_ham} is strange in the context of electric circuits, it is ubiquitous mechanically, particularly in rotational systems. Examples of two such rotational systems are shown in figure \ref{fig:cos_pot}

For the pendulum, the potential energy is gravitational, and we can find,
\begin{equation}
    U = m g L (1 - \cos{(\theta)}) 
\end{equation}\label{eqn:pend_pot}
Implying $E_j \to m g L$. For the slider spring system, with the spring unstretched at $\theta = 0$,
\begin{align}
    \begin{split}
        U =& \frac{1}{2} k \left( L \sin{\left( \frac{\theta}{2}\right)}\right)^2 \\
        =& \frac{1}{4} k L^2 (1 - \cos{(\theta)})
    \end{split}
\end{align}
Implying $E_j \to \frac{1}{4} k L^2$.

\section{Capacitively Shunted Josephson Junction}

We may use the mechanical systems outlined in the previous section to find a complete mechanical analog to the capacitively shunted Josephson junction. The quantum hamiltonian for the circuit shown in figure \ref{fig:cap_s_jj} a) is given by,
\begin{equation}\label{eqn:transmon_h}
    H = 4 E_c (n-n_g)^2  + E_j (1 - \cos{\left(\varphi - \varphi_{ext}\right)})
\end{equation}
Where $\varphi = \frac{\phi}{\Phi_0}$ is the normalized flux, and $n$ is a conjugate variable representing the number of cooper pairs (each with charge $2e$) on the capacitor plates,
\begin{equation}
    n=\frac{q}{2e} = - i \frac{\partial}{\partial \varphi}
\end{equation}
$E_C=\frac{e^2}{2C}$ represents the capacitance in energy units. Here $n_g$ has been introduced, representing a static offset charge on the capacitor plates, which will always be present in a real device, a source of noise that was common in the early days of superconducting qubits.

\begin{figure}
    \centering
    \includegraphics[width=\textwidth]{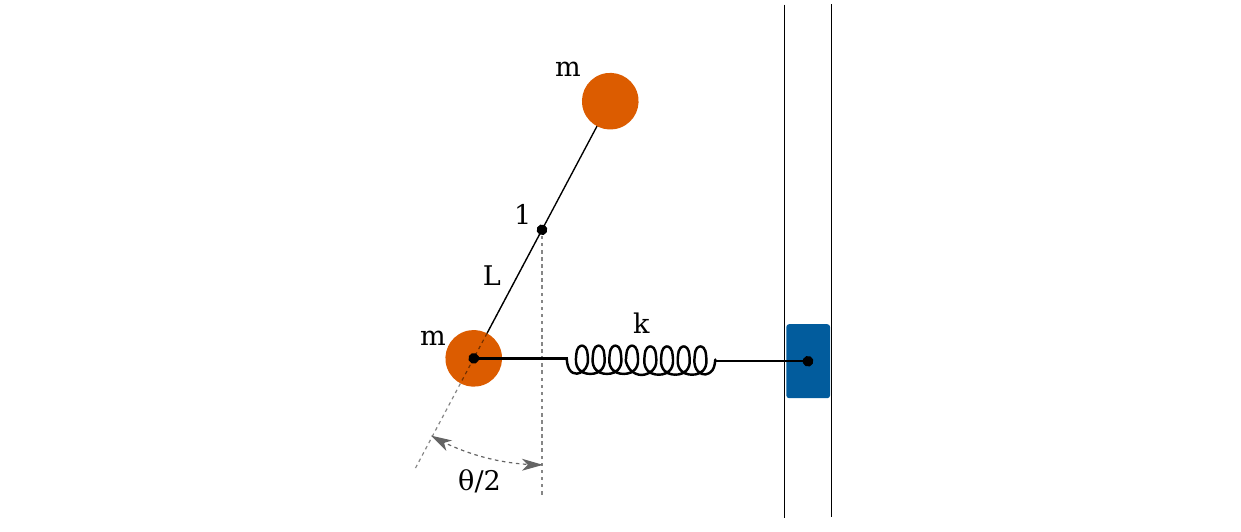}
    \caption{\textbf{A mechanical analog to the Capacitively shunted Josephson Junction} A balanced pendulum is coupled to a slider spring. The pendulum, which has total length $2L$, is pinned in the middle at (1), and rotates freely. The moment of inertia of the pendulum is given by $I = 2 m L^2$. The slider spring setup is identical to that described in figure \ref{fig:transmon}. }
    \label{fig:transmon}
\end{figure}

The balanced pendulum and slider spring system shown in figure \ref{fig:transmon} is an analagous mechanical system.

The Lagrangian of this system can be found as,
\begin{align}
\begin{split}
    L =& \frac{1}{2} I \left( \frac{\Dot{\theta}}{2}\right)^2 - \frac{1}{4} k L^2 \left( 1 - \cos{\left( \theta\right)}\right)
\end{split}
\end{align}
Defining the conjugate momentum in the standard way,
\begin{equation}
    p = \frac{1}{4} I \Dot{\theta}
\end{equation}
We can then find the Hamiltonian as,
\begin{equation}
    H = \frac{1}{2 (I/4)} p^2 + \frac{1}{4} k L^2 (1 - \cos{(\theta)})
\end{equation}
Where $\{p, \theta\} =0$ and $[\theta, p] = i \hbar$. The mapping from electrical to mechanical parameters is then clear,
\begin{align}
    \begin{split}
        E_C \to & \frac{\hbar^2}{2I} \\
        E_j \to & \frac{1}{4} k L^2 \\
        \varphi \to & \theta
    \end{split}
\end{align}
We can therefore think of increasing $E_C$ as being equivalent to decreasing $I$ and increasing $E_j$ as equivalent to increasing $k$, which are independent changes. If unbalanced pendulum analogy had been used (as in figure \ref{fig:cos_pot} a), increasing $E_j$ independently of $E_c$ would correspond to increasing $g$, which is less intuitive than making a spring stiffer. For this reason, the slider spring analogy will be preferentially used in this work.

The mechanism shown in figure \ref{fig:transmon} is invaluable for understanding the reasoning behind the Transmon qubit \cite{Koch2007}, which was developed to overcome sensitivity to the offset charge $n_g$, as in equation \ref{eqn:transmon_h}. Namely, the Transmon operates in a parameter regime $E_j/E_C$ such that the first few energy levels of the Hamiltonian are nearly completely insensitive to $n_g$. To see how this is done, consider the eigenvalue problem,

\begin{equation}\label{eqn:transmon_evp}
    E\psi(\varphi) = 4 E_c (-i \frac{\partial}{\partial \varphi}-n_g)^2 \psi(\varphi)  + E_j (1 - \cos{\left(\varphi - \varphi_{ext}\right)})\psi(\varphi)
\end{equation}

Since the potential is $2 \pi$ periodic in $\varphi$, the boundary condition $\psi(\varphi) = \psi(\varphi + 2 \pi)$ applies. This boundary condition is also clear from the mechanical analogy, rotating the pendulum in figure \ref{fig:transmon} through $\pi$ will land it in an indistinguishable state from where it started. We can make the substitution $\psi(\varphi) \to e^{i n_g \varphi} g\left( \frac{\varphi}{2}\right)$ to remove the $n_g$ dependence from \ref{eqn:transmon_evp},

\begin{equation}\label{eqn:transmon_evp}
    (E - E_j)  g\left( \frac{\varphi}{2}\right) = - E_c \frac{\partial^2 g\left( \frac{\varphi}{2}\right)}{\partial \varphi^2}  - E_j\cos{\left(\varphi - \varphi_{ext}\right)})g\left( \frac{\varphi}{2}\right)
\end{equation}

Which now has the boundary condition $g\left( \frac{\varphi}{2}\right) = e^{2 \pi i n_g} g\left( \frac{\varphi}{2} + \pi\right)$. From this we can see that the offset charge only effects the solution when it "wraps around" the boundary. Therefore, if we confine the oscillations to small values of $\theta$, we would expect the eigenvalues to be insensitive to charge. This corresponds to using a very stiff spring $k$ for a given pendulum intertia $I$, or in electrical terms making the ratio $E_j/E_c$ large. This is precisely the transmon regime, which can be rigorously identified using asymptotics on Mathieu's characteristic values. From the pendulum the downside of the Transmon design is also clear: if the pendulum is confined to small $\theta$ the nonlinearity will also be weak. Using the small angle approximation we would expect the nonlinearity to decrease with $\theta^2$, which is indeed the case.

\section{Fluxonium}

\begin{figure}
    \centering
    \includegraphics[width=\textwidth]{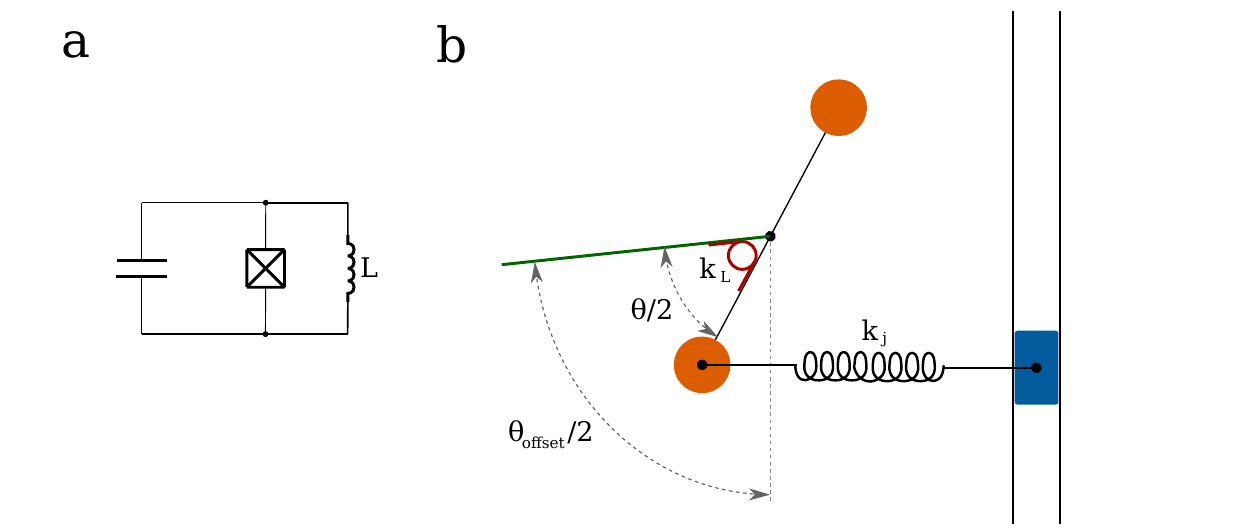}
    \caption{\textbf{The Fluxonium qubit a)} The fluxonium circuit, which is just a capacitively shunted Josephson junction in parallel with a linear inductance $L$. \textbf{b)} An analogy to the fluxonium may be found by adding a linear torsion spring (red) to the mechanism discussed in figure \ref{fig:transmon} . The torsion spring is fixed to the pendulum and an adjustable, stationary rod (green). This spring has rate $k_L$, while the slider spring has rate $k_j$. The angle $\theta_{\text{offset}}$ has also been introduced to allow for an analogy to external flux. Moving the green rod is equivalent to tuning the external flux.  Note that the definition of the angle $\theta/2$ has also changed, such that the angle seen by the slider spring mechanism is $\frac{\theta - \theta_{\text{offset}}}{2}$}
    \label{fig:mech_fluxonium}
\end{figure}

The fluxonium is slightly contemporary to the Transmon \cite{Manucharyan_2009}, and attempts to solve the offset charge problem by introducing a large shunting inductance to the circuit given in Fig.~\ref{fig:cap_s_jj}, as shown in Fig.~\ref{fig:mech_fluxonium} a. The resulting electrical hamiltonian is,
\begin{equation}
    H = 4 E_c n^2 + \frac{1}{2} E_L \varphi^2 + E_j (1 - \cos{\left(\varphi - \varphi_{ext}\right)})
\end{equation}
Where $E_l = \frac{\Phi_0^2}{L}$. Note that the appropriate boundary conditions for the associated Schrodinger equation are $\psi(\pm\infty) = 0$, and the solutions should be entirely insensitive to offset charge, as can be seen via the same procedure used in the last section.

Adding a torsion spring to the mechanism devised previously results in a mechanical analogy to the fluxonium, as shown in figure \ref{fig:mech_fluxonium} b.

The Hamiltonian of the mechanical system can be found as,
\begin{equation}
    H = \frac{1}{2 (I/4)} p^2 + \frac{1}{8} k_l \theta^2 + \frac{1}{4} k_j l^2 (1 - \cos{(\theta - \theta_{\text{offset}})})
\end{equation}
Where $2l$ now represents the pendulum length. The mapping from electrical to mechanical parameters is therefore,

\begin{align}
    \begin{split}
        E_C \to & \frac{\hbar^2}{2I} \\
        E_j \to & \frac{1}{4} k_j l^2 \\
        E_L \to & \frac{1}{8} k_L  \\
        \varphi \to & \theta \\
        \varphi_{\text{ext}} \to & \theta_{\text{offset}}
    \end{split}
\end{align}

The fluxonium operates in the regime $\frac{E_j}{E_L} \propto \frac{k_j}{k_L} \gg 1$. From the mechanical analogy, it is clear that this is required to get any interesting behavior, as if the torsion spring was much stiffer than the slider spring the mechanism would behave harmonically, oscillating about $\theta = 0$. The behavior of the mechanism is clearly also sensitive to $\theta_{\text{offset}}$. In the limit $\frac{k_j}{k_L} \gg 1$, the two lowest energy statitically stable configurations of the mechanism will be when the pendulum is oriented almost vertically, and $\theta \approxeq \theta_{\text{offset}}$ or $\theta \approxeq \theta_{\text{offset}} - 2 \pi$. Therefore, we would expect the ground state and first excited states should correspond to anharmonic oscillations around these stable points. This is desirable, as having ground and first excited state wavefunctions localized in entirely separate regions of space (disjoint) corresponds to protection from noise, namely decay from the first excited state to the ground state. The energy difference between these states $E_{10}$ corresponds to the difference in the potential energy stored in the torsion spring, which should increase with $|\pi - \theta_{\text{offset}}|$. To first order in $\frac{k_L}{k_j}$, static analysis of the mechanism yields,
\begin{equation}\label{eqn:classical_fluxonium_e10}
    E_{10} \approxeq 2 \pi k_L|\theta_{\text{offset}} - \pi|
\end{equation}

\begin{figure}
    \centering
    \includegraphics[width=\textwidth]{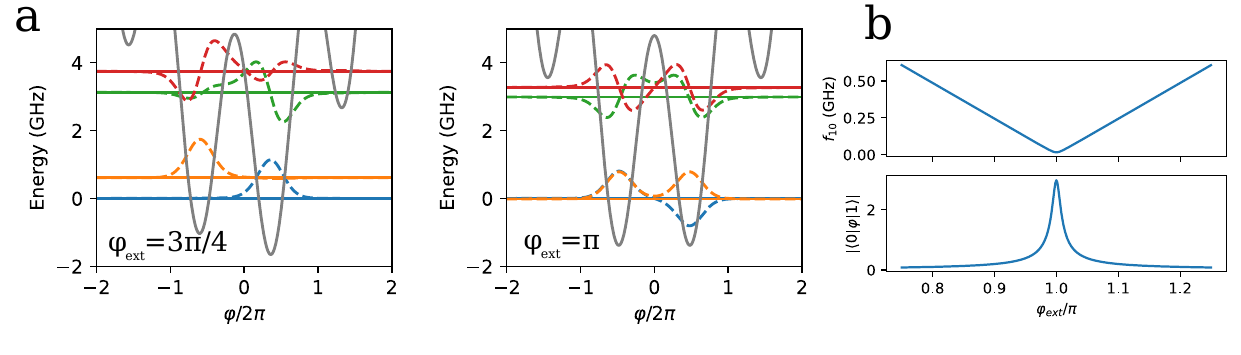}
    \caption{\textbf{Solutions to the Fluxonium Hamiltonian a)} The 4 lowest energy states of fluxonium qubits with $\varphi_{\text{ext}} = 3 \pi/4$ and $\pi$, respectively, found via numerical simulation. The solid lines indicate eigenenergies, while the dashed lines of the same color show the associated eigenfunction. The grey line shows the potential. We can see that when $\varphi_{\text{ext}}$ is far from $\pi$ the two lowest energy states $\ket{0}$ and $\ket{1}$ states are localized in separate wells of the cosine potential and are far from degeneracy. The next highest states $\ket{2}$ and $\ket{3}$ are far detuned from the low-lying states. When $\varphi_{\text{ext}}$ is tuned to $\pi$ the $\ket{0}$ and $\ket{1}$ states become nearly degenerate, forming symmetric and anti-symmetric combinations of the ground states of each well. \textbf{b)} The qubit frequency $f_{10}$ and flux matrix element vs $\varphi_{\text{ext}}$. Following intuition, the qubit frequency approaches zero as the external flux approaches $\pi$. The flux matrix element is nearly zero everywhere except for near degeneracy.    }
    \label{fig:fluxonium_wfs}
\end{figure}
Classically, if $\theta_{\text{offset}} = \pi$ the two states should be indistinguishable and the energy gap should be exactly 0. This would correspond to degenerate energy levels, which we would expect to be broken in the quantum case by a tunneling process, forming wavefunctions consisting of symmetric and anti-symmetric combinations of oscillations about each stable point.

The degeneracy of the low energy states also suggests that $\varphi_{\text{ext}} = \pi$ should be a flux insensitive point for the fluxonium, since quantum mechanically the energy difference $E_{10}$ should be a smooth function of $\varphi_{\text{ext}}$, resembling an avoided crossing. Therefore, if $E_{10}$ is minimized at $\varphi_{\text{ext}} = \pi$, it must also be true that at that point $\frac{\partial E_{10}}{\partial \varphi_{\text{ext}}} = 0$. However, equation \ref{eqn:classical_fluxonium_e10} shows that the classical solution is not smooth, $\frac{\partial E_{10}}{\partial \theta_{\text{offset}}}$ is singluar at $\theta_{\text{offset}} = \pi$. Therefore, any flux insensitivty found near $\varphi_{\text{ext}}$ in the quantum case is purely a result of the smoothing of $E_{10}$ due to tunnelling, and we should not expect that the fluxonium will simultaneously have disjoint qubit states and protection from noise in $\varphi_{\text{ext}}$. In fact, this indicates that we should never expect that any one dimensional mechanism could simultaniously have disjoint and nearly degenerate eigenstates. It seems one dimension simply does not provide enough room for both of these properties to exist at the same time.

Fig.~\ref{fig:fluxonium_wfs} shows results of numerically solving the fluxonium schrodinger equation for a particular set of parameter values. We see that the mechanical analogy gave us accurate qualitive information about the behavior of the low-lying eigenstates. Namely, as shown in Fig.~\ref{fig:fluxonium_wfs} a, the ground state and first excited state are located in separate wells for $\varphi_{\text{ext}} < \pi$, and form the aforementioned symmetric and anti-symmetric superpositions at $\varphi = \pi$. Fig.~\ref{fig:fluxonium_wfs} b shows $f_{10} = \frac{E_{10}}{2 \pi \hbar}$ and the flux matrix element between the ground and first excited state $|\bra{0}\varphi\ket{1}|$ as a function of $\varphi_{\text{ext}}$. Mechanical analysis predicted well the linear nature of $f_{10}$ away from degeneracy. 

Despite the fact that the fluxonium is not formally protected from decay of the excited state at near-degeneracy, qubits of very high quality have been constructed to operate at this point. These qubits take advantage of the flux insenitivity at this point and that the power of the noise that leads to decay from the excited state tends to decrease quadratically with transition frequency. Fluxoniums with coherence times of over $1ms$ and gate times on the order of $10ns$ have been constructed \cite{Somoroff2021}, which implies that on the order of $10^5$ single qubit gates could be completed in a coherence time, approximately $50$ times more than what the qubits that achieved quantum supremacy could do \cite{Arute2019}.

\section{$0-\pi$ Qubit}

The $0-\pi$ qubit is a type of protected superconducting qubit. The design originates from ideas of encoding qubits "topologically" in arrays of Josephson junctions \cite{Ioffe2002, Kitaev2006}. Practically, these ideas have been distilled down to the circuit shown in figure \ref{fig:cap_s_jj}, which is the simplest version of the $0-\pi$. This circuit may be reduced to the Hamiltonian \cite{Groszkowski2018},
\begin{align}
    \begin{split}
        H =& E_{C\phi} n_{\phi}^2 +  E_{C\theta} n_{\theta}^2 - 2 E_j \cos{(\theta)}\cos{(\phi - \frac{\varphi_{\text{ext}}}{2})} + E_l \phi^2\\
        =& E_{C\phi} n_{\phi}^2 +  E_{C\theta} n_{\theta}^2 - E_j \left( \cos{\left(\theta + \phi - \frac{\varphi_{\text{ext}}}{2}  \right)} + \cos{\left(\theta - \phi + \frac{\varphi_{\text{ext}}}{2}  \right)}\right)
    \end{split}
\end{align}
Where the second line is found from the first using a trigonometric identity. This is a 2D Hamiltonian, despite the fact that the full circuit has 3 degrees of freedom. This occurs in the ideal limit when all corresponding circuit elements are identical, leading to a separable Hamiltonian and a decoupling of a harmonic mode. 

\begin{figure}
    \centering
    \includegraphics[width=\textwidth]{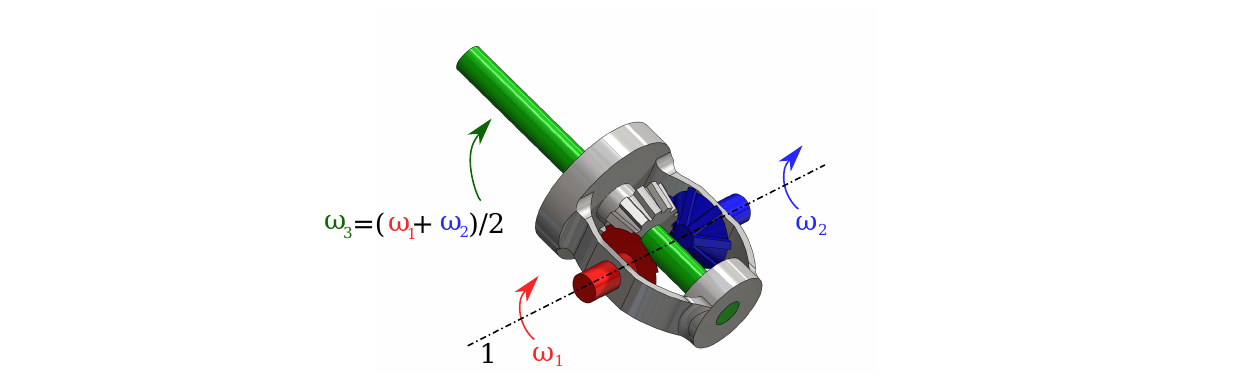}
    \caption{\textbf{The spur-gear differential} The spur-gear differential has two input shafts (red and blue) and one output shaft (green). The angular velocity of the ouput shaft $\omega_3$ is the average of the angular velocities of the input shafts, $\omega_3 = (\omega_1 + \omega_2)/2.$ }
    \label{fig:differential}
\end{figure}

The sum of cosines form of the Hamiltonian is particularly useful when trying to find a mechanical analogy to the $0-\pi$ Hamiltonian. Adding and subtracting angles is a natural task for planetary gearboxes. In particular, the differential gearbox, a common automotive component, is perfectly suited for the task. A spur-gear differential is shown in figure \ref{fig:differential}.

\begin{figure}
    \centering
    \includegraphics[width=\textwidth]{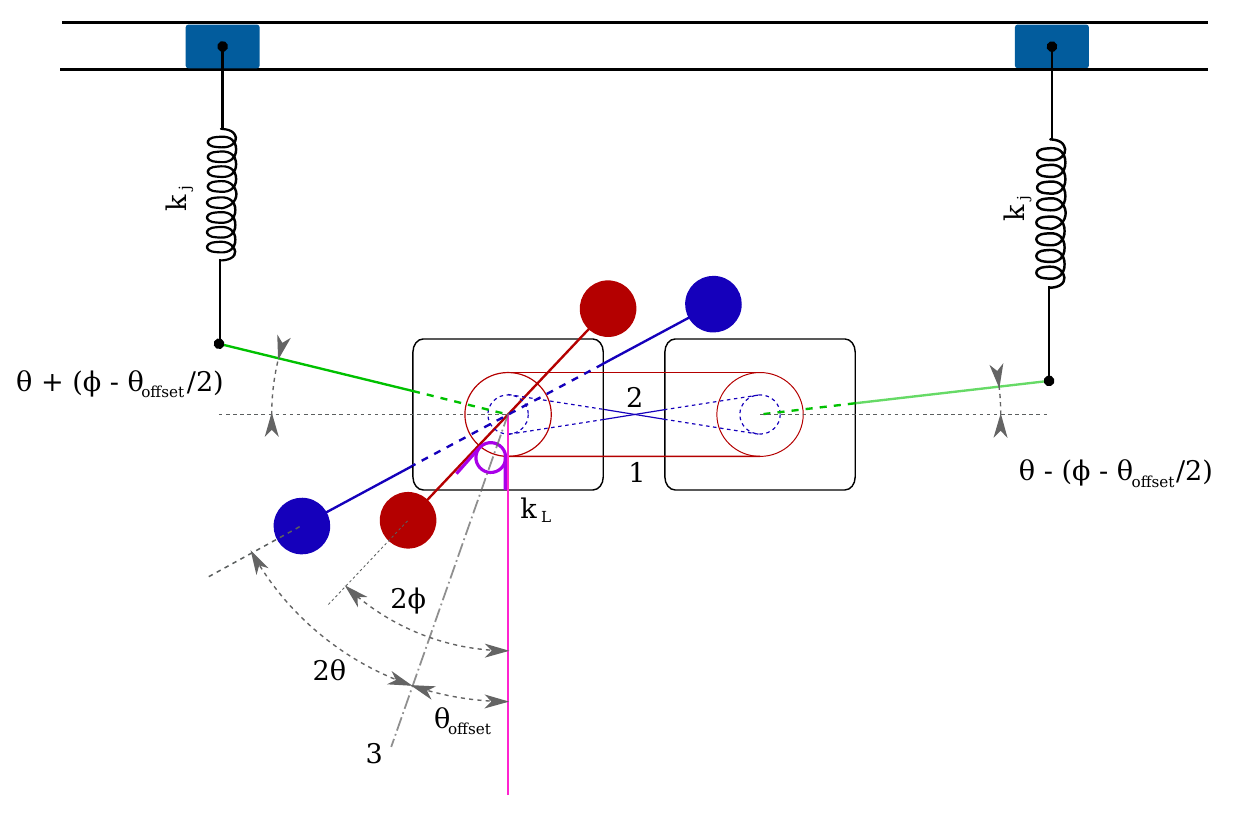}
    \caption{\textbf{A mechanical analogy to the full zero-pi qubit Hamiltonian} The $0-\pi$ qubit is analagous to a mechanism consisting of two pendula, two slider springs, and two differential gearboxes, represented by grey boxes in the diagram. The inputs and outputs to the gearboxes are color-coded identically to figure \ref{fig:differential}. Therefore, we can see that the inputs to one of the gearboxes are the angles of the red and blue pendula, which are coupled to the inputs of the second gearbox by belts (1 and 2). One of the belts has a twist (2), which reverses the direction of rotation. This allows one of the gearboxes to add the angles of the pendula, while the other subtracts them. The construction line (3) indicates the pendula angles at which the slider springs are unstretched, $\theta = 0$ and $\phi = \frac{\theta_{\text{offset}}}{2}$. }
    \label{fig:zero_pi}
\end{figure}

Using this gearbox, it is straightforward to assemble the parts we have used previously into a $0-\pi$ qubit. The result is shown in figure \ref{fig:zero_pi}.
The Hamiltonian of this mechanism is given by,
\begin{equation}
    H = \frac{1}{2(4 I_{\phi})} p_{\phi}^2 + \frac{1}{2(4 I_{\theta})} p_{\theta}^2 + \frac{1}{2}(4 k_L) \phi^2 - \frac{1}{2} k_j L^2 \left(\cos{\left(\theta + \phi - \frac{\theta_{\text{offset}}}{2}\right)} + \cos{\left(\theta - \phi + \frac{\theta_{\text{offset}}}{2}\right)}\right) 
\end{equation}
Where $\theta$ and $\phi$ have been re-used to represent the angles analagous to the same fluxes. Here, $L$ is the length of the output arm of the gearboxes. The mapping between electrical and mechanical parameters is then,
\begin{align}
    \begin{split}
        E_{C\phi} \to& \frac{\hbar^2}{2(4 I_{\phi})}\\
        E_{C\theta} \to& \frac{\hbar^2}{2(4 I_{\theta})}\\
        E_l \to& \frac{1}{2}(4 k_L)\\
        E_j \to& \frac{1}{2} k_j L^2\\
        \varphi_{\text{ext}} \to & \theta_{\text{offset}}\\
    \end{split}
\end{align}
Similar to the fluxonium, the $0-\pi$ qubit typically operates at $\varphi_{\text{ext}} = 0$, which is a degenerate, flux insensitive point. Since the $0-\pi$ is two dimensional, it is able to simultaneously have degenerate, disjoint eigenstates. This is possible because choosing the right parameter values leads to tunnelling happening in only one dimension, such that eigenstates may remain disjoint in the other \cite{Dempster2014}. 

Understanding this parameter range and property in terms of the mechanism shown in figure \ref{fig:zero_pi} remains an unsolved problem, that will be the subject of further study. Is it a coincidence that "topologically" protected qubits map to mechanisms with rich kinematic constraints such as planetary gearboxes? Or are the two connected in an important way? Answering this question may open up new avenues in qubit design.

\printbibliography

@article{odom06,
  title = {New Measurement of the Electron Magnetic Moment Using a One-Electron Quantum Cyclotron},
  author = {Odom, B. and Hanneke, D. and D'Urso, B. and Gabrielse, G.},
  journal = {Phys. Rev. Lett.},
  volume = {97},
  issue = {3},
  pages = {030801},
  numpages = {4},
  year = {2006},
  month = {Jul},
  publisher = {American Physical Society},
  doi = {10.1103/PhysRevLett.97.030801},
  url = {https://link.aps.org/doi/10.1103/PhysRevLett.97.030801}
}

@ARTICLE{shannon48,
  author={Shannon, C. E.},
  journal={The Bell System Technical Journal}, 
  title={A mathematical theory of communication}, 
  year={1948},
  volume={27},
  number={3},
  pages={379-423},
  doi={10.1002/j.1538-7305.1948.tb01338.x}}

@article{feynman1982simulating,
  author = {Feynman, Richard P},
  journal = {International journal of theoretical physics},
  number = {6/7},
  pages = {467--488},
  publisher = {World Scientific},
  title = {Simulating physics with computers},
  volume = 21,
  year = 1982
}

@article{Shor_1997,
	doi = {10.1137/s0097539795293172},
  
	url = {https://doi.org/10.1137%2Fs0097539795293172},
  
	year = 1997,
	month = {oct},
  
	publisher = {Society for Industrial {\&} Applied Mathematics ({SIAM})},
  
	volume = {26},
  
	number = {5},
  
	pages = {1484--1509},
  
	author = {Peter W. Shor},
  
	title = {Polynomial-Time Algorithms for Prime Factorization and Discrete Logarithms on a Quantum Computer},
  
	journal = {{SIAM} Journal on Computing}
}

@book{nielsen00,
  author = {Nielsen, Michael A. and Chuang, Isaac L.},
  publisher = {Cambridge University Press},
  title = {Quantum Computation and Quantum Information},
  year = 2000
}

@article{Fowler_2012,
	doi = {10.1103/physreva.86.032324},
  
	url = {https://doi.org/10.1103%2Fphysreva.86.032324},
  
	year = 2012,
	month = {sep},
  
	publisher = {American Physical Society ({APS})},
  
	volume = {86},
  
	number = {3},
  
	author = {Austin G. Fowler and Matteo Mariantoni and John M. Martinis and Andrew N. Cleland},
  
	title = {Surface codes: Towards practical large-scale quantum computation},
  
	journal = {Physical Review A}
}

@article{Calderbank_1996,
	doi = {10.1103/physreva.54.1098},
  
	url = {https://doi.org/10.1103%2Fphysreva.54.1098},
  
	year = 1996,
	month = {aug},
  
	publisher = {American Physical Society ({APS})},
  
	volume = {54},
  
	number = {2},
  
	pages = {1098--1105},
  
	author = {A. R. Calderbank and Peter W. Shor},
  
	title = {Good quantum error-correcting codes exist},
  
	journal = {Physical Review A}
}

@article{Ruster_2016,
	doi = {10.1007/s00340-016-6527-4},
  
	url = {https://doi.org/10.1007%2Fs00340-016-6527-4},
  
	year = 2016,
	month = {sep},
  
	publisher = {Springer Science and Business Media {LLC}
},
  
	volume = {122},
  
	number = {10},
  
	author = {T. Ruster and C. T. Schmiegelow and H. Kaufmann and C. Warschburger and F. Schmidt-Kaler and U. G. Poschinger},
  
	title = {A long-lived Zeeman trapped-ion qubit},
  
	journal = {Applied Physics B}
}

@article{Vool2017,
archivePrefix = {arXiv},
arxivId = {1610.03438},
author = {Vool, Uri and Devoret, Michel},
doi = {10.1002/cta.2359},
eprint = {1610.03438},
issn = {1097007X},
journal = {International Journal of Circuit Theory and Applications},
number = {7},
pages = {897--934},
title = {{Introduction to quantum electromagnetic circuits}},
volume = {45},
year = {2017}
}

@article{shnirman97,
  title = {Quantum Manipulations of Small Josephson Junctions},
  author = {Shnirman, Alexander and Sch\"on, Gerd and Hermon, Ziv},
  journal = {Phys. Rev. Lett.},
  volume = {79},
  issue = {12},
  pages = {2371--2374},
  numpages = {0},
  year = {1997},
  month = {Sep},
  publisher = {American Physical Society},
  doi = {10.1103/PhysRevLett.79.2371},
  url = {https://link.aps.org/doi/10.1103/PhysRevLett.79.2371}
}

@article{Brooks2013,
archivePrefix = {arXiv},
arxivId = {1302.4122},
author = {Brooks, Peter and Kitaev, Alexei and Preskill, John},
doi = {10.1103/PhysRevA.87.052306},
eprint = {1302.4122},
issn = {10502947},
journal = {Physical Review A - Atomic, Molecular, and Optical Physics},
number = {5},
title = {{Protected gates for superconducting qubits}},
volume = {87},
year = {2013}
}

@article{Koch2007,
archivePrefix = {arXiv},
arxivId = {cond-mat/0703002},
author = {Koch, Jens and Yu, Terri M. and Gambetta, Jay and Houck, A. A. and Schuster, D. I. and Majer, J. and Blais, Alexandre and Devoret, M. H. and Girvin, S. M. and Schoelkopf, R. J.},
doi = {10.1103/PhysRevA.76.042319},
eprint = {0703002},
issn = {10502947},
journal = {Physical Review A - Atomic, Molecular, and Optical Physics},
number = {4},
pages = {1--21},
primaryClass = {cond-mat},
title = {{Charge-insensitive qubit design derived from the Cooper pair box}},
volume = {76},
year = {2007}
}

@article{Ioffe2002,
archivePrefix = {arXiv},
arxivId = {cond-mat/0205186},
author = {Ioffe, L. B. and Feigel'man, M. V.},
doi = {10.1103/PhysRevB.66.224503},
eprint = {0205186},
issn = {1550235X},
journal = {Physical Review B - Condensed Matter and Materials Physics},
number = {22},
pages = {1--8},
primaryClass = {cond-mat},
title = {{Possible realization of an ideal quantum computer in Josephson junction array}},
volume = {66},
year = {2002}
}

@article{Kitaev2006,
archivePrefix = {arXiv},
arxivId = {cond-mat/0609441},
author = {Kitaev, Alexei},
eprint = {0609441},
pages = {1--6},
primaryClass = {cond-mat},
title = {{Protected qubit based on a superconducting current mirror}},
url = {http://arxiv.org/abs/cond-mat/0609441},
year = {2006}
}

@article{Groszkowski2018,
archivePrefix = {arXiv},
arxivId = {1708.02886},
author = {Groszkowski, Peter and Paolo, A. Di and Grimsmo, A. L. and Blais, A. and Schuster, D. I. and Houck, A. A. and Koch, Jens},
doi = {10.1088/1367-2630/aab7cd},
eprint = {1708.02886},
issn = {13672630},
journal = {New Journal of Physics},
number = {4},
pages = {1--21},
title = {{Coherence properties of the 0-$\pi$ qubit}},
volume = {20},
year = {2018}
}

@article{Burkard2005,
archivePrefix = {arXiv},
arxivId = {cond-mat/0408588},
author = {Burkard, Guido},
doi = {10.1103/PhysRevB.71.144511},
eprint = {0408588},
issn = {10980121},
journal = {Physical Review B - Condensed Matter and Materials Physics},
number = {14},
pages = {1--8},
primaryClass = {cond-mat},
title = {{Circuit theory for decoherence in superconducting charge qubits}},
volume = {71},
year = {2005}
}

@article{Manucharyan_2009,
	doi = {10.1126/science.1175552},
  
	url = {https://doi.org/10.1126%2Fscience.1175552},
  
	year = 2009,
	month = {oct},
  
	publisher = {American Association for the Advancement of Science ({AAAS})},
  
	volume = {326},
  
	number = {5949},
  
	pages = {113--116},
  
	author = {Vladimir E. Manucharyan and Jens Koch and Leonid I. Glazman and Michel H. Devoret},
  
	title = {Fluxonium: Single Cooper-Pair Circuit Free of Charge Offsets},
  
	journal = {Science}
}

@article{Somoroff2021,
archivePrefix = {arXiv},
arxivId = {2103.08578},
author = {Somoroff, Aaron and Ficheux, Quentin and Mencia, Raymond A. and Xiong, Haonan and Kuzmin, Roman V. and Manucharyan, Vladimir E.},
eprint = {2103.08578},
title = {{Millisecond coherence in a superconducting qubit}},
url = {http://arxiv.org/abs/2103.08578},
volume = {1},
year = {2021}
}

@article{Arute2019,
author = {Arute, Frank and Arya, Kunal and Babbush, Ryan and Bacon, Dave and Bardin, Joseph C. and Barends, Rami and Biswas, Rupak and Boixo, Sergio and Brandao, Fernando G.S.L. and Buell, David A. and Burkett, Brian and Chen, Yu and Chen, Zijun and Chiaro, Ben and Collins, Roberto and Courtney, William and Dunsworth, Andrew and Farhi, Edward and Foxen, Brooks and Fowler, Austin and Gidney, Craig and Giustina, Marissa and Graff, Rob and Guerin, Keith and Habegger, Steve and Harrigan, Matthew P. and Hartmann, Michael J. and Ho, Alan and Hoffmann, Markus and Huang, Trent and Humble, Travis S. and Isakov, Sergei V. and Jeffrey, Evan and Jiang, Zhang and Kafri, Dvir and Kechedzhi, Kostyantyn and Kelly, Julian and Klimov, Paul V. and Knysh, Sergey and Korotkov, Alexander and Kostritsa, Fedor and Landhuis, David and Lindmark, Mike and Lucero, Erik and Lyakh, Dmitry and Mandr{\`{a}}, Salvatore and McClean, Jarrod R. and McEwen, Matthew and Megrant, Anthony and Mi, Xiao and Michielsen, Kristel and Mohseni, Masoud and Mutus, Josh and Naaman, Ofer and Neeley, Matthew and Neill, Charles and Niu, Murphy Yuezhen and Ostby, Eric and Petukhov, Andre and Platt, John C. and Quintana, Chris and Rieffel, Eleanor G. and Roushan, Pedram and Rubin, Nicholas C. and Sank, Daniel and Satzinger, Kevin J. and Smelyanskiy, Vadim and Sung, Kevin J. and Trevithick, Matthew D. and Vainsencher, Amit and Villalonga, Benjamin and White, Theodore and Yao, Z. Jamie and Yeh, Ping and Zalcman, Adam and Neven, Hartmut and Martinis, John M.},
doi = {10.1038/s41586-019-1666-5},
isbn = {4158601916},
issn = {14764687},
journal = {Nature},
number = {7779},
pages = {505--510},
pmid = {31645734},
publisher = {Springer US},
title = {{Quantum supremacy using a programmable superconducting processor}},
url = {http://dx.doi.org/10.1038/s41586-019-1666-5},
volume = {574},
year = {2019}
}

@article{Dempster2014,
archivePrefix = {arXiv},
arxivId = {1402.7310},
author = {Dempster, Joshua M. and Fu, Bo and Ferguson, David G. and Schuster, D. I. and Koch, Jens},
doi = {10.1103/PhysRevB.90.094518},
eprint = {1402.7310},
issn = {1550235X},
journal = {Physical Review B - Condensed Matter and Materials Physics},
number = {9},
pages = {1--12},
title = {{Understanding degenerate ground states of a protected quantum circuit in the presence of disorder}},
volume = {90},
year = {2014}
}

\end{document}